\documentclass[aps,prl,twocolumn,groupedaddress,showpacs]{revtex4}
\usepackage{graphicx,color}
\def\vc#1{\mbox{\boldmath $#1$}}

\begin{document}


\title{One-dimensional $\alpha$ condensation of $\alpha$-linear-chain states in ${^{12}{\rm C}}$ and ${^{16}{\rm O}}$}


\author{T.~\textsc{Suhara}$^1$, Y.~\textsc{Funaki}$^2$, B.~\textsc{Zhou}$^3$, H.~\textsc{Horiuchi}$^4$, and A.~\textsc{Tohsaki}$^4$}
\affiliation{$^1$Matsue College of Technology, Matsue 690-8518, Japan}
\affiliation{$^2$Nishina Center for Accelerator-Based Science, The Institute of Physical and Chemical Research (RIKEN), Wako 351-0198, Japan}
\affiliation{$^3$Department of Physics, Nanjing University, Nanjing 210093, China}
\affiliation{$^4$Research Center for Nuclear Physics (RCNP), Osaka University, Osaka 567-0047, Japan}


\date{\today}

\begin{abstract}
We present a new picture that the $\alpha$-linear-chain structure for ${^{12}{\rm C}}$ and ${^{16}{\rm O}}$ has one-dimensional $\alpha$ condensate character.
The wave functions of linear-chain states which are described by superposing a large number of Brink wave functions have extremely large overlaps of nearly 100\% 
with single Tohsaki-Horiuchi-Schuck-R\"opke (THSR) wave functions, which were proposed to describe the $\alpha$ condensed``gas-like'' states.
Although this new picture is different from the conventional idea of the spatial localization of  $\alpha$ clusters,
the density distributions are shown to have localized $\alpha$-clusters which is due to the inter-$\alpha$ Pauli repulsion.
\end{abstract}



\maketitle


The cluster formation is one of the fundamental degrees of freedom in nuclear structure~\cite{tang}. 
In particular, for the last decade, the physics of $\alpha$-particle condensation has triggered much interest~\cite{thsr,concept}. 
The best achievement in this direction is to have revealed that the Hoyle state (the $0^+_2$ state at $7.67$ MeV in ${^{12}{\rm C}}$) does not only have the $3\alpha$-cluster structure but also has an $\alpha$-condensate-like structure of the $3\alpha$ clusters, which occupy an identical $S$ orbit of mean-field potential~\cite{funaki_2003}. 
In ${^{16}{\rm O}}$ the analog to the Hoyle state was also proposed to exist theoretically as the $0^+_6$ state, which might be identified as the observed $15.1$ MeV state~\cite{4aocm}. 

The above $\alpha$ condensate character was clarified with the use of the condensate-type microscopic cluster model wave function, so-called Tohsaki-Horiuchi-Schuck-R\"opke (THSR) wave function~\cite{thsr,funaki_2002}. 
The wave function does not put $\alpha$ clusters on spatial positions any more like localized clustering but generates a quantum orbit occupied by the clusters, where they move in a non-localized way in a mean-field type potential of the $\alpha$ clusters. 
The THSR wave function has been used for the analysis of the ``gas-like'' state.
The ``gas-like'' state is here defined to be the dilute cluster state whose density is a fraction of the normal density, where clusters are interacting weakly with each other, although this ``gas-like'' object is still much denser than an ordinary gas.
For the ``gas-like'' states such as ${^8{\rm Be}}$ and the Hoyle state, the solutions of 2$\alpha$ and 3$\alpha$ resonating group method (RGM) equation under the bound state approximation were found to be almost 100\% equivalent to single 2$\alpha$ and 3$\alpha$ THSR wave functions, respectively~\cite{concept}.

Quite recently, in Refs.~\cite{bo1,bo2}, the THSR-type $\alpha+{^{16}{\rm O}}$ cluster model wave function was applied to the investigation of the inversion doublet bands of $\alpha+{^{16}{\rm O}}$ cluster structure in ${^{20}{\rm Ne}}$.
In spite of the fact that the existence of the inversion doublet bands proves most convincingly the localized clustering of $\alpha+{^{16}{\rm O}}$~\cite{horiuchi_20Ne}, they showed that the single THSR wave functions are almost 100\% equivalent to the solutions of $\alpha+{^{16}{\rm O}}$ RGM equation.
Since the THSR wave function expresses non-localized clustering, this result requires us to change the traditional understanding of clustering as the localized cluster structure. 
We should also note that for the ground state of ${^{12}{\rm C}}$ the THSR wave function coincides by 93\% with the 3$\alpha$ RGM wave function~\cite{concept}.

Another typical example of the localized clustering is the $\alpha$-linear-chain state, where $\alpha$ clusters are aligned on a line. 
Thus the above mentioned results urge us to examine the non-localized picture of the THSR ansatz for $\alpha$-linear-chain states that $\alpha$ clusters occupy an identical lowest orbit of a prolate-deformed mean-field potential. 

The 3$\alpha$-linear-chain state, which was first proposed by Morinaga about 60 years ago, was used to explain a structure of the Hoyle state~\cite{morinaga}. 
Although this idea for the Hoyle state was later replaced by the interpretation as a ``gas-like'' state of $3\alpha$ clusters, the higher $0^+$ state observed at $10.3$ MeV has still been the possible candidate of the chain state, for which antisymmetrized molecular dynamics (AMD) and fermionic molecular dynamics (FMD) calculations predict a $3\alpha$ structure with slight bending from the linear-chain arrangement~\cite{fmd,amd}.

As for the $4\alpha$-linear-chain states in ${^{16}{\rm O}}$, Chevallier {\it et al.} observed the resonant $2^+, 4^+$, and $6^+$ states in an excitation energy region of $17 \sim 20$ MeV in a reaction ${^{12}{\rm C}}(\alpha,{^{8}{\rm Be}}){^{8}{\rm Be}}$, which lie on a $J(J+1)$ rotational trajectory with $\hbar^2/2{\cal I}=64$ keV, consistent with a linear-chain of $4\alpha$ particles~\cite{chevallier}. 
Theoretical calculation of $\alpha$ reduced widths supported this picture~\cite{suzuki_chain}. 
Freer {\it et al.} also observed those states, which lie on the $J(J+1)$ rotational trajectory with $\hbar^2/2{\cal I}=95$ keV, still indicating extremely deformed shape~\cite{freer_4alpha1,freer_4alpha2}. 
On the other hand, at higher angular-momentum region, recently Ichikawa {\it et al.} reported the formation and the stability of 4$\alpha$-linear-chain state based on the cranked Hartree-Fock method which does not assume the existence of $\alpha$ clusters.
They showed that rotating 4$\alpha$-linear-chain states with $13\hbar \sim 18\hbar$ are formed and stable~\cite{ichikawa}. 

In addition to these examples, various theoretical and experimental studies of $N\alpha$-linear-chain states have been made for almost half a century~\cite{freer_review1,freer_review2}. 
However, even now, the understanding is insufficient. 
To clarify the nature of linear-chain states is very important from historical view.
In this Letter, we compare the solutions of 3$\alpha$- and 4$\alpha$-linear-chain states obtained by the generator coordinate method (GCM) by using Brink wave functions with the THSR wave functions. 
From this comparison, we present a new understanding of the linear-chain structure.

First let us define the localized $n\alpha$ cluster wave function, which is known as the Brink wave function~\cite{brink}:
\begin{equation}
\Phi^{(B)}_J(\vc{R}_1,\cdots,\vc{R}_n)={\hat P}^J_{M0}{\hat P}^+ {\cal A}[\varphi^\alpha_1 \cdots  \varphi^\alpha_n],\label{eq:brink1}
\end{equation}
\begin{equation}
\varphi^\alpha_i =\phi_{\small \vc{R}_i}\phi_{\small \vc{R}_i}\phi_{\small \vc{R}_i}\phi_{\small \vc{R}_i}\chi_{p\uparrow}\chi_{p\downarrow}\chi_{n\uparrow}\chi_{n\downarrow},\label{eq:brink2}
\end{equation}
\begin{equation}
\phi_{\small \vc{R}_i} =(\pi b^2)^{-3/4}\exp \Big[-\frac{1}{2b^2} (\vc{r}-\vc{R}_i)^2 \Big],\label{eq:brink3}
\end{equation}
where $\vc{R}_i$ is the position parameter of the $i$-th $\alpha$ particle, and ${\hat P}^J_{M0}$ and ${\hat P}^+$ are the projection operators onto the angular momentum $JM$ and positive parity, respectively. In all the subsequent calculations, we impose $(\vc{R}_1+\cdots + \vc{R}_n)/n=\vc{0}$ for the choice of the position-parameter values, to eliminate the spurious center-of-mass motion.

For the $N N$ interaction, the Volkov No.2 
force~\cite{volkov} is used with Majorana parameter $M=0.59$ and $M=0.63$ for ${^{12}{\rm C}}$ and ${^{16}{\rm O}}$, respectively. For these choices of the Majorana parameter, the binding energies of the ground states of ${^{12}{\rm C}}$ and ${^{16}{\rm O}}$ are calculated to be $-91.8$ MeV and $-129.9$ MeV, respectively, which are in good agreement with the corresponding experimental data, $-92.2$ MeV and $-127.6$ MeV. As the size parameter of the $\alpha$ particle, we adopt $b=1.376$ fm in Eq.~(\ref{eq:brink3}), which gives a minimum energy of the $\alpha$ particle, $-28.0$ MeV.

We consider model space of the pure linear-chain state, by taking $R_{ix}=R_{iy}=0$. 
In this model space we perform the GCM calculation by taking $\vc{R}_i=(0,0,R_{iz})$ as discretized generator coordinates.
The GCM wave function can thus be written as,
\begin{equation}
\Phi_J^{\rm (B-GCM)}=\sum_{R_{1z},\cdots,R_{nz}}f(R_{1z},\cdots,R_{nz})\Phi_J^{\rm (B)}(R_{1z},\cdots,R_{nz}). \label{gcmwf}
\end{equation}
The coefficients $f(R_{1z},\cdots,R_{nz})$ can then be determined by solving the Hill-Wheeler equation,
\begin{eqnarray}
&&\hspace{-0.5cm}\sum_{R_{1z}^\prime, \cdots, R_{nz}^\prime} \langle \Phi^{(B)}_J(R_{1z},\cdots,R_{nz}) |\hat{H}-E | \Phi^{(B)}_J(R_{1z}^\prime,\cdots,R_{nz}^\prime) \rangle \nonumber \\
&& \hspace{2cm} \times f(R_{1z}^\prime,\cdots,R_{nz}^\prime) =0. \label{eq:hweq}
\end{eqnarray}
In order to obtain the converged solution of the above equation, we adopt $100$ and $300$ parameter values for $R_{1z},\cdots,R_{nz}$.
These are determined by randomly generated distances between the adjacent clusters,
$d_{i} \equiv R_{(i+1)z} -  R_{iz}$, within a range of $2\ {\rm fm} \le d_{i} \le 10\ {\rm fm}$.
The optimal parameter set is selected, on the variational principle, out of 8 sets of randomly generated 100 or 300 parameter values.

\begin{figure}[tb]
\begin{center}
\includegraphics[width=8.6cm]{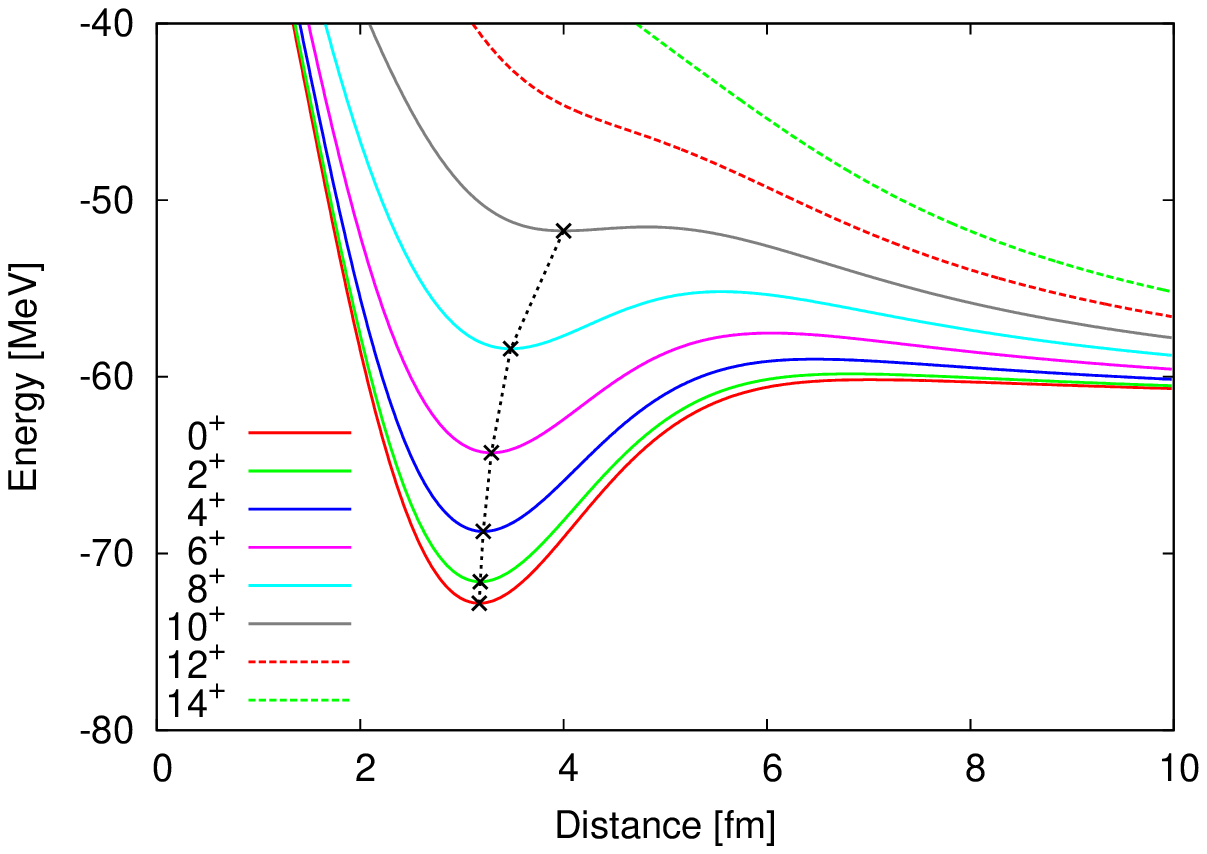}
\includegraphics[width=8.6cm]{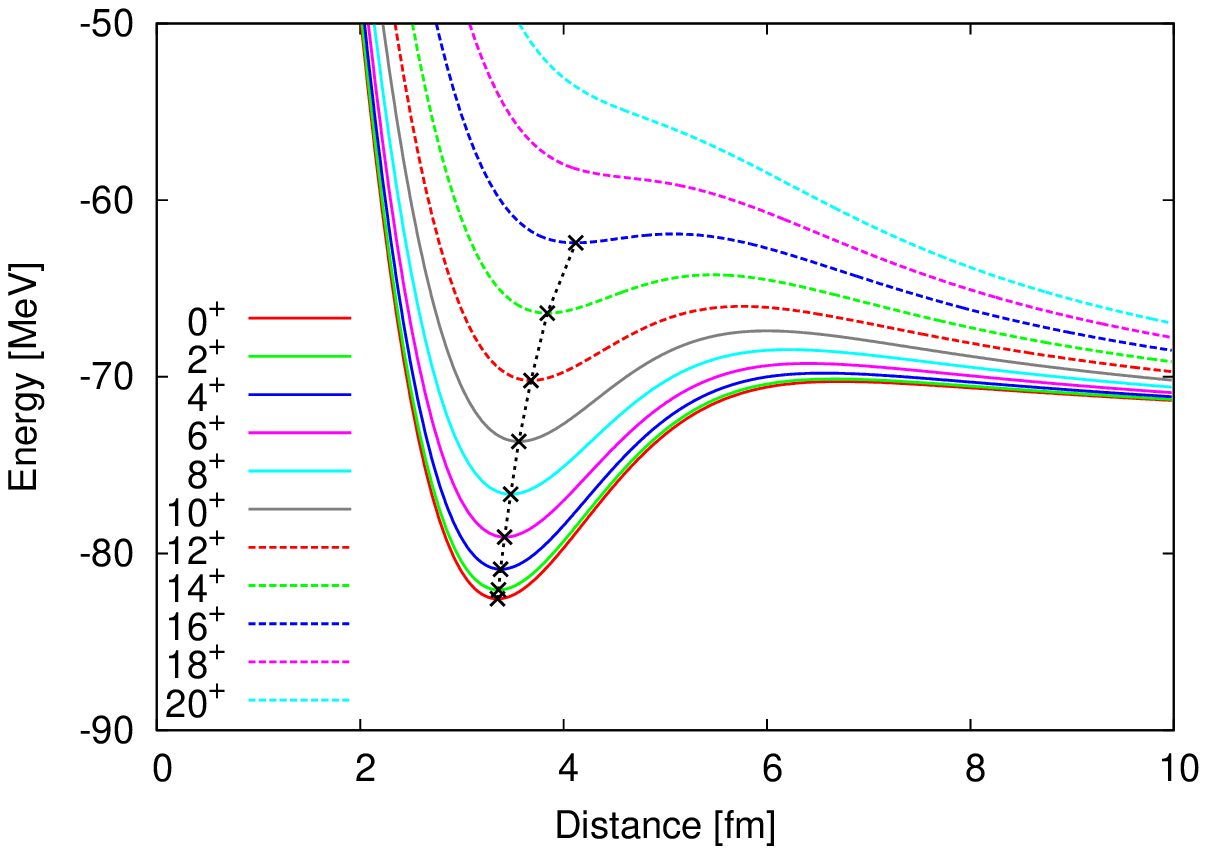}
\caption{The energy curves for $J^{\pi}$ states in equidistant linear-chain configuration,
$d_1=d_2$ for ${^{12}{\rm C}}$ (Top) and $d_1=d_2=d_3$ for ${^{16}{\rm O}}$ (Bottom). 
The minima are denoted by crosses.}
\label{fig:12}
\end{center}
\end{figure}

First we show the energy curves for ${^{12}{\rm C}}$ in the top panel of Fig.~\ref{fig:12}, where an equal distance between $\alpha$ particles is assumed, i.e. $d_1=d_2$. 
For the states with $J^\pi=0^+,2^+,\cdots,10^+$, the minimum-energy states appear, which are denoted by crosses. 
The distances $d_1=d_2$ which give the minimum energies are pushed outside as the increase of the angular momentum. 
This indicates that the centrifugal force, which is strengthened by the increase of the angular momentum, is leading to a more elongated linear-chain structure.
The minima tend to disappear at higher angular momentum. 
Similar calculation is performed for ${^{16}{\rm O}}$, which is shown in the bottom panel of Fig.~\ref{fig:12}, where $d_1=d_2=d_3$ is assumed. 
The minimum-energy states survive up to around $J^\pi=16^+$. 
We should note that the higher angular-momentum states correspond to the states with $J \simeq 13 \sim 18$ obtained by the cranked Hartree-Fock calculation in Ref.~\cite{ichikawa}, which have been shown to be stable against the bending motion of the $\alpha$ particles from the linear-chain arrangement. 

\begin{table}[tb]
\begin{center}
\caption{Minimum energies given by intrinsic (denoted as Int.), parity-projected (denoted as Pos.), angular-momentum-projected, and GCM wave functions of Brink model, in a unit of MeV. 
The distance parameter values $(d_1,d_2)$ for ${^{12}{\rm C}}$ and $(d_1, d_2, d_3)$ for ${^{16}{\rm O}}$, respectively, which give the minimum energies, are also shown, in a unit of fm. 
The threshold energies for ${^{12}{\rm C}}$ and ${^{16}{\rm O}}$ are calculated to be $E_{3\alpha}^{th}=-83.97$ MeV and $E_{4\alpha}^{th}=-111.96$ MeV, respectively.}\label{tab:1}
\begin{tabular}{cccclccc}
\hline\hline
 & \multicolumn{3}{c}{${^{12}{\rm C}}$ $(3\alpha)$} &  & \multicolumn{3}{c}{${^{16}{\rm O}}$ $(4\alpha)$} \\
\hline
 & $(d_1,d_2)$ & Energy & GCM &  & $(d_1,d_2,d_3)$ & Energy & GCM \\
\hline
Int. & $(3.0,3.0)$ & $-63.7$ &  &  & $(3.3,3.3,3.3)$ & $-72.3$ &  \\
Pos. & $(2.3,3.8)$ & $-64.4$ &  &  & $(4.1,3.3,2.8)$ & $-73.8$ &  \\
$0^+$ & $(2.5,4.1)$ & $-73.9$ & $-76.8$ &  & $(4.2,3.4,2.8)$ & $-84.2$ & $-90.1$ \\
$2^+$ & $(2.5,4.1)$ & $-72.7$ & $-75.8$ &  & $(4.2,3.4,2.8)$ & $-83.7$ & $-89.7$ \\
$4^+$ & $(2.6,4.2)$ & $-70.0$ & $-73.6$ &  & $(4.2,3.4,2.9)$ & $-82.6$ & $-88.9$ \\
\hline\hline
\end{tabular}
\end{center}
\end{table}

In Table~\ref{tab:1}, we list the energy minima, which appeared for intrinsic states (Int.), for the states after parity projection (Pos.) and angular-momentum projection ($J^\pi$), and the lowest eigenenergies by the GCM calculation (GCM). 
The energy of $0^+$ state obtained by GCM is $-76.8$ MeV, $7.2$ MeV above the $3\alpha$ threshold energy, which is still about $4$ MeV higher than the energy of $0^+$ state observed at $10.3$ MeV, for which AMD and FMD calculations predict a $3\alpha$ structure with slight bending from the linear-chain arrangement.
This difference may come from the lack of degree of freedom of bending motion of the $\alpha$ cluster from the linear-chain configuration in our present model space.

Next, we show the THSR wave function as follows: 
\begin{eqnarray}
&& \Phi_{J}^{\rm (THSR)}(\beta_{x}, \beta_{y}, \beta_{z}) = {\hat P}^J_{M0} \Phi^{\rm (THSR)}(\beta_{x}, \beta_{y}, \beta_{z}) = \nonumber \\
&& {\hat P}^J_{M0} {\cal A}\ \Big\{ \prod_{i=1}^n\exp \Big[- 
\sum_{k=x,y,z}\frac{2}{B_k^2} (X_{ik}-X_{Gk})^2 \Big] \phi(\alpha_i) \Big\},\nonumber \\ 
 \label{eq:thsr} 
\end{eqnarray}
with the antisymmetrizer $\cal A$ operating on all nucleons and $B_k^2=b^2+2\beta_k^2$. $\vc{X}_i$ and $\vc{X}_G$ are the center-of-mass coordinate of the $i$-th $\alpha$ particle and the total center-of-mass coordinate, respectively. $\phi(\alpha_i)$ is the 
intrinsic wave function of the $i$-th $\alpha$ particle as follows:
\begin{equation}
\phi(\alpha_i) \propto \exp\Big[-\sum_{1\leq k<l \leq4}({\vc r}_{i,k} - 
{\vc r}_{i,l})^2/(8b^2)\Big]\chi_{p\uparrow}\chi_{p\downarrow}\chi_{n\uparrow}\chi_{n\downarrow}.\label{eq:int_alpha}
\end{equation}
The only parameter in this wave function is $\vc{B}$, or equivalently $(\beta_{x}, \beta_{y}, \beta_{z})$, which is the size parameter of mean-field potential of the $\alpha$ clusters and corresponds to the size of a whole nucleus. 

\begin{figure}[tb]
\begin{center}
\includegraphics[width = 8.6cm]{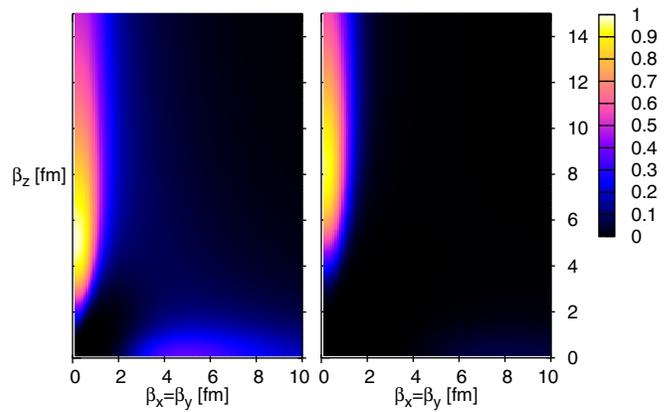}
\caption{The squared overlap surfaces between the linear-chain wave functions obtained by GCM and the single THSR wave functions in two parameter spaces, $\beta_x=\beta_y$ and $\beta_z$, for $J^\pi=0^+$ states in ${^{12}{\rm C}}$ (Left) and ${^{16}{\rm O}}$ (Right).}
\label{fig:34}
\end{center}
\end{figure}

In the THSR model space, a linear-chain structure is realized by taking strongly prolate-deformed values for the parameter $(\beta_{x}, \beta_{y}, \beta_{z})$, i.e. $\beta_x=\beta_y\sim 0$ and non-zero $\beta_z$ value.
This has the one-dimensional $\alpha$ condensate structure, in which the $\alpha$ clusters move independently inside a mean-field potential with strongly prolate-deformed shape.
This THSR-type picture therefore is very different from the traditional one for the linear-chain structure, in which localized clustering in one dimension has been considered to occur.
Our aim of this paper is to show that the solutions of the linear-chain Brink-GCM equation are quite close to single THSR-type linear-chain wave functions.

In Fig.~\ref{fig:34}, we show the squared overlap surfaces between the Brink-GCM wave functions of Eq.~(\ref{gcmwf}) and the THSR wave functions Eq.~(\ref{eq:thsr}) for $J^\pi=0^+$ states of ${^{12}{\rm C}}$ and ${^{16}{\rm O}}$, respectively, in two parameter spaces, $\beta_x=\beta_y$ and $\beta_z$. 
Surprisingly, we can see very large maximum values for both cases. 
The region of having large values is widely ranged in the extremely prolate-deformed region. 
We list the maximum values in Table~\ref{tab:2}, for $J^\pi=0^+,2^+,4^+$ states in ${^{12}{\rm C}}$ and ${^{16}{\rm O}}$, together with the corresponding $(\beta_{x} = \beta_{y}, \beta_{z})$ values to give the maxima.
The energies of the THSR wave functions are also shown.
The maximum squared overlaps are more than $0.98$ and $0.93$ in all the $J^\pi$ states listed for ${^{12}{\rm C}}$ and ${^{16}{\rm O}}$, respectively. 
The energies of the THSR wave functions are very close, with a few hundred keV, to those of the GCM wave functions listed in Table~\ref{tab:1}.
These results of course mean that the THSR wave function very well describes the pure linear-chain structure. 
Considering the fact that 100 and 300 bases are necessary to reach the convergence in solving the Hill-Wheeler equation Eq.~(\ref{eq:hweq}) for ${^{12}{\rm C}}$ and ${^{16}{\rm O}}$, respectively, it is surprising that the single THSR wave functions coincide with such a large number of superposed Brink wave functions with very high accuracy for both cases. 
This obviously urges us to reconsider the conventional picture of linear-chain structure. 
The one-dimensional ``gas'' of the $\alpha$ particles with the condensate character is very different from the localized clustering described by a single Brink wave function. 
The maximum squared overlaps between the GCM wave functions and single Brink wave functions are only 78\% and 48\% for the $0^+$ states of ${^{12}{\rm C}}$ and ${^{16}{\rm O}}$, respectively.
These values are significantly smaller than the squared overlaps, 99\% and 94\%, between the GCM wave functions and THSR wave fucntions.

\begin{table}[tb]
\begin{center}
\caption{Maximum squared overlaps between the GCM wave functions obtained based on the Brink wave function and the single THSR wave functions, for $J^\pi=0^+,2^+,4^+$ states for ${^{12}{\rm C}}$ and ${^{16}{\rm O}}$. 
$(\beta_x=\beta_y,\beta_z)$ giving maxima and their energies are denoted, in units of fm and MeV, respectively.}\label{tab:2}
\begin{tabular}{ccccccc}
\hline\hline
 & \multicolumn{3}{c}{${^{12}{\rm C}}$ $(3\alpha)$} & \multicolumn{3}{c}{${^{16}{\rm O}}$ $(4\alpha)$} \\
\hline
 & $(\beta_x=\beta_y,\beta_z)$ & Max. & Energy & $(\beta_x=\beta_y,\beta_z)$ & Max. & Energy \\
\hline
$0^+$ & $(0.1,5.1)$ & $0.987$ & $-76.6$ & $(0.1,8.2)$ & $0.944$ & $-89.8$ \\
$2^+$ & $(0.1,5.4)$ & $0.989$ & $-75.6$ & $(0.1,8.4)$ & $0.942$ & $-89.4$ \\
$4^+$ & $(0.1,6.6)$ & $0.981$ & $-73.4$ & $(0.1,9.0)$ & $0.931$ & $-88.8$ \\
\hline\hline
\end{tabular}
\end{center}
\end{table}

\begin{figure}[tb]
\begin{center}
\includegraphics[angle=270,scale=0.7]{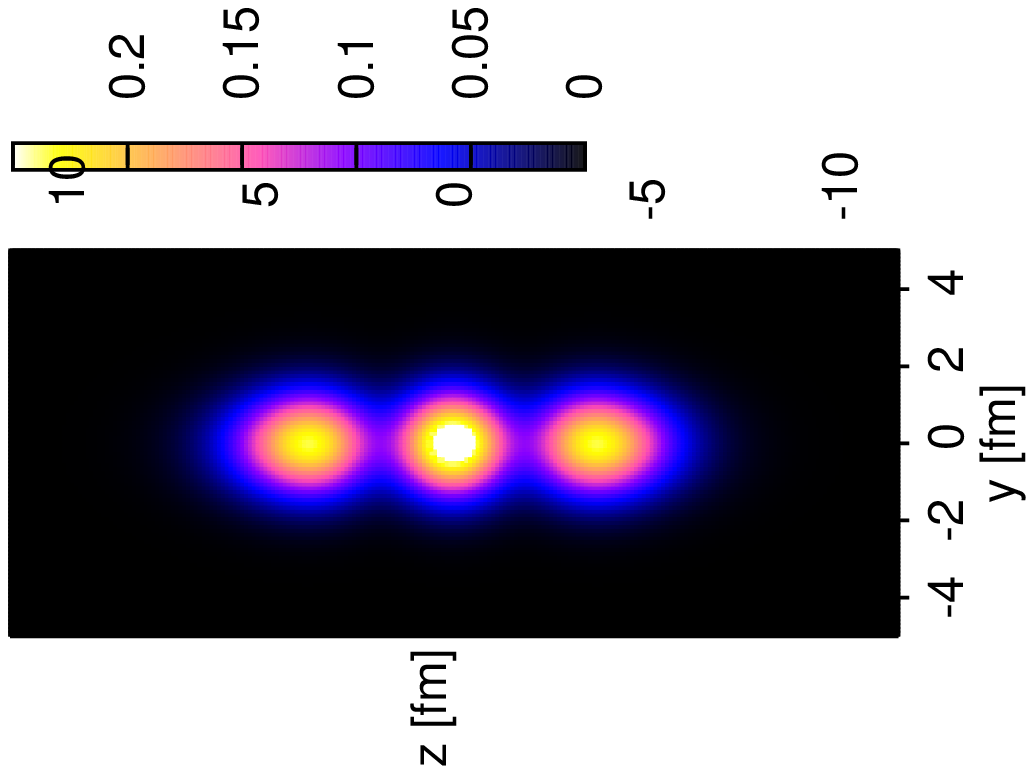}
\includegraphics[angle=270,scale=0.7]{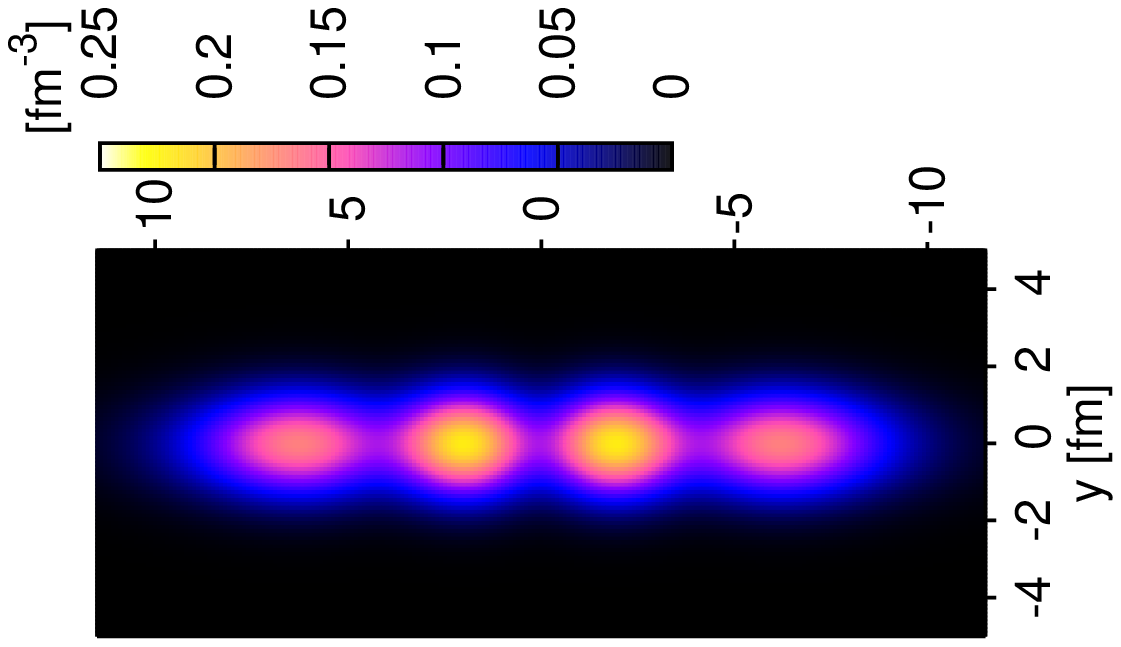}
\caption{Intrinsic density profiles of the $3\alpha$- (Left) and $4\alpha$- (Right) linear-chain states generated from the THSR wave functions before angular-momentum projection at ($\beta_x=\beta_y=0.1\ {\rm fm}$, $\beta_z=5.1\ {\rm fm}$) and ($\beta_x=\beta_y=0.1\ {\rm fm}$, $\beta_z=8.2\ {\rm fm}$), respectively.
}
\label{fig:5}
\end{center}
\end{figure}

We show in Fig.~\ref{fig:5} the intrinsic density profiles of THSR wave functions,
\begin{equation}
\rho(\vc{r}) = \frac{\langle \Phi^{\rm (THSR)} | \sum_{i} \delta(\vc{r}-\hat{\vc{r}}_{i}-\vc{X}_G) | \Phi^{\rm (THSR)} \rangle}{\langle \Phi^{\rm (THSR)} | \Phi^{\rm (THSR)} \rangle},
\end{equation}
at $x = 0$ for ${^{12}{\rm C}}$ ($\beta_x=\beta_y=0.1$ fm, $\beta_z=5.1$ fm) and ${^{16}{\rm O}}$ ($\beta_x=\beta_y=0.1$ fm, $\beta_z=8.2$ fm), which give the maximum squared overlaps for $J^\pi=0^+$ states, respectively.
We can clearly see the $3\alpha$ and $4\alpha$ clusters, which are aligned on a straight line and are all localized.
This result gives us a new understanding of clustering.
Even for the states which are described by non-localized-type wave function,
localized nature of clustering can appear in density distribution due to the Pauli principle.
We can say that dynamics prefers non-localized clustering but kinematics coming from the Pauli principle makes the system look like localized clustering.
This seems to be a common feature for the microscopic wave function having a shape with extremely prolate deformation, in which the cluster motions are restricted in one dimension. 
The ``gas-like'' feature is in particular expressed as the long tail extended in the $z$ direction.

In summary, we investigated the $3\alpha$- and $4\alpha$-linear-chain states with the use of Brink wave functions, where the $\alpha$-particle motion is restricted to one dimension along $z$-axis. 
We obtained energy minima for a wide range of angular-momentum states for both cases. 
We obtained the Brink-GCM wave functions with sufficient energy convergence by superposing 100 and 300 Brink wave functions adopted as bases to solve the Hill-Wheeler equation, for ${^{12}{\rm C}}$ and ${^{16}{\rm O}}$, respectively. 
Then we compared the Brink-GCM wave functions with the single THSR wave functions, which provide the $\alpha$ condensate structure. 
The maximum squared overlaps between them amount to more than 98\% for ${^{12}{\rm C}}$ and 93\% for ${^{16}{\rm O}}$, for $J^\pi=0^+, 2^+, 4^+$ states, at the shapes with strongly prolate deformation. 
These large overlap values at the extremely deformed shapes mean that the $\alpha$-linear-chain states have the one-dimensional $\alpha$ condensate character, where the $\alpha$ clusters are trapped into a one-dimensional potential in a non-localized manner, like a ``gas''. 
Although this new picture is different from the conventional idea of the spatial localization of  $\alpha$ clusters, the density distributions are shown to have localized $\alpha$-clusters which is due to the inter-$\alpha$ Pauli repulsion.
To clarify the magnitude of the Pauli effect in linear-chain states,
we are investigating, for example, momentum distribution of $\alpha$ clusters in the $z$-direction.

We comment on the bent linear-chain state. 
There is a possibility that real linear-chain structure bends.
However, to understand the real linear-chain structure, it is very useful to know that one-dimensional condensate character is more essential than the localization character for the ideal straight linear-chain states.

The non-localized clustering picture was introduced in the study of inversion doublet of ${^{20}{\rm Ne}}$~\cite{bo1,bo2}, which had been the important basis of the localized 2-body clustering picture. 
The present work has shown that the non-localized picture plays a crucial role in the more important example of the localized many-body clustering, the $\alpha$-linear-chain state.
This picture thus seems to hold in general cluster systems, and then opens a new horizon for the nuclear cluster physics.

\section*{Acknowledgments}
The authors wish to thank P. Schuck, G. R\"opke, T. Yamada, Z. Z. Ren, and C. Xu for many helpful discussions concerning this work. 
This work was supported by JSPS KAKENHI Grant Numbers 25400256, 25400288, and 25887049.


\begin{thebibliography}{99}
\bibitem{tang}
K. Wildermuth and Y. C. Tang, {\it A Unified Theory of the Nucleus} (Vieweg, Braunschweig, 1977).
\bibitem{thsr}
A. Tohsaki, H. Horiuchi, P. Schuck, and G. R\"opke, Phys. Rev. Lett. {\bf 87}, 192501 (2001).
\bibitem{concept}
Y. Funaki, H. Horiuchi, W. von Oertzen, G. R\"{o}pke, P. Schuck, A. Tohsaki, and T. Yamada, 
Phys. Rev. C {\bf 80}, 064326 (2009).
\bibitem{funaki_2003}
Y. Funaki, A. Tohsaki, H. Horiuchi, P. Schuck, and G. R\"opke, Phys. Rev. C {\bf 67}, 051306(R) (2003).
\bibitem{4aocm}
Y. Funaki, T. Yamada, H. Horiuchi, G. R\"opke, P. Schuck, and A. Tohsaki, 
Phys. Rev. Lett. {\bf 101}, 082502 (2008).
\bibitem{funaki_2002}
Y. Funaki, H. Horiuchi, A. Tohsaki, P. Schuck, and G. R\"opke, Prog. Theor. Phys. {\bf 108}, 297 (2002).
\bibitem{bo1}
B. Zhou, Z. Z. Ren, C. Xu, Y. Funaki, T. Yamada, A. Tohsaki, H. Horiuchi, P. Schuck, and G. R\"opke,
Phys. Rev. C {\bf 86}, 014301 (2012).
\bibitem{bo2}
B. Zhou, Y. Funaki, H. Horiuchi, Z. Z. Ren, G. R\"opke, P. Schuck, A. Tohsaki, C. Xu, and T. Yamada,
Phys. Rev. Lett. {\bf 110}, 262501 (2013).
\bibitem{horiuchi_20Ne}
H. Horiuchi and K. Ikeda, Prog. Theor. Phys. {\bf 40}, 277 (1968).
\bibitem{morinaga}
H. Morinaga, Phys. Rev. {\bf 101}, 254 (1956); Phys. Lett. {\bf 21}, 78 (1966).
\bibitem{fmd}
T. Neff and H. Feldmeier, Nucl. Phys. A {\bf 738}, 357 (2004).
\bibitem{amd}
Y. Kanada-En'yo, Prog. Theor. Phys. {\bf 117}, 655 (2007).
\bibitem{chevallier}
P. Chevallier, F. Scheibling, G. Goldring, I. Plesser, and M. W. Sachs, Phys. Rev. {\bf 160}, 827 (1967).
\bibitem{suzuki_chain}
Y. Suzuki, H. Horiuchi, and K. Ikeda, Prog. Theor. Phys. {\bf 47}, 1517 (1972).
\bibitem{freer_4alpha1}
M. Freer {\it et al.}, Phys. Rev. C {\bf 51}, 1682 (1995).
\bibitem{freer_4alpha2}
M. Freer {\it et al.}, Phys. Rev. C {\bf 70}, 064311 (2004).
\bibitem{ichikawa}
T. Ichikawa, J. A. Maruhn, N. Itagaki, and S. Ohkubo, Phys. Rev. Lett. {\bf 107}, 112501 (2011).
\bibitem{freer_review1}
M. Freer and A. C. Merchant, J. Phys. G: Nucl. Part. Phys. {\bf 23}, 261 (1997).
\bibitem{freer_review2}
M. Freer, Rep. Prog. Phys. {\bf 70}, 2149 (2007).
\bibitem{brink}
D. M. Brink, {\it Proc. Int. School of Physics Enrico Ferm, Course 36}, Varenna, ed. C. Bloch (Academic Press, New York, 1966).
\bibitem{volkov}
A. B. Volkov, Nucl. Phys. A {\bf 74}, 33 (1965). 
\end{thebibliography}
\end{document}